\documentclass[]{style/ceurart}
\usepackage{listings}
\usepackage{xcolor}
\lstdefinestyle{simplehmstyle}{
    basicstyle=\ttfamily\small,  
    breaklines=true,             
    showstringspaces=false,      
    tabsize=2,                   
    frame=single,
    keywordstyle=\bfseries,
    morekeywords={
        String, Int, Float, Data,
        Message, Argument,        
        DebateRequest, RetrievalResponse, SystemResponse,
        UserTurn, Simulation,
        EvaluationRequest, EvaluationResponse,
    },
    commentstyle=\color{green!50!black}, 
    morecomment=[l]//,                   
    morecomment=[l]\#,                   
}
\begin{document}
\copyrightyear{2025}
\copyrightclause{Copyright for this paper by its authors. Use permitted under Creative Commons License Attribution 4.0 International (CC BY 4.0).}
\conference{CLEF 2025 Working Notes, 9 -- 12 September 2025, Madrid, Spain}

\title{DS@GT at Touch\'e: Large Language Models for Retrieval-Augmented Debate}
\title[mode=sub]{Notebook for the Touch\'e Lab at CLEF 2025}

\author[1]{Anthony Miyaguchi}[
orcid=0000-0002-9165-8718,
email=acmiyaguchi@gatech.edu,
]
\cormark[1]

\author[1]{Conor Johnston}[
orcid=0009-0001-1777-9255,
email=cjohnston@gatech.edu,
]
\cormark[1]

\author[1]{Aaryan Potdar}[
email=apotdar31@gatech.edu,
]
\cormark[1]

\address[1]{Georgia Institute of Technology, North Ave NW, Atlanta, GA 30332}
\cortext[1]{Corresponding author.}

\begin{abstract}
    Large Language Models (LLMs) demonstrate strong conversational abilities. 
    In this Working Paper, we study them in the context of debating in two ways: their ability to perform in a structured debate along with a dataset of arguments to use and their ability to evaluate utterances throughout the debate.
    We deploy six leading publicly available models from three providers for the Retrieval-Augmented Debate and Evaluation.
    The evaluation is performed by measuring four key metrics: Quality, Quantity, Manner, and Relation. 
    Throughout this task, we found that although LLMs perform well in debates when given related arguments, they tend to be verbose in responses yet consistent in evaluation.
    The accompanying source code for this paper is located at \url{https://github.com/dsgt-arc/touche-2025-rad}.
\end{abstract}

\begin{keywords}
  Large Language Models (LLM) \sep
  Retrieval-Augmented Generation (RAG) \sep
  Argumentative Systems \sep
  Self-Evaluation \sep
  Conversational AI \sep
  CEUR-WS \sep
  DS@GT
\end{keywords}

\maketitle
\section{Introduction}

The DS@GT team participated in the Retrieval-Augmented Debating (RAD) task as part of the Touché 2025 competition \cite{kiesel:2025}.
The RAD task involves generating responses to a simulated debate partner and evaluating the systems that generate these responses.
In this paper, we describe our approach to the competition that leverages retrieval-augmented generation (RAG) using the ClaimRev corpus and large language models (LLMs).
We design an automatic evaluation pipeline to assess our responses across the required dimensions in a scalable and consistent way.
\section{Related Work}

LLMs are effective debaters \cite{khan2024debatingpersuasivellmsleads}.
Although not examined in a debate environment, LLMs enhanced with Retrieval Augmented Generation (RAG) systems have been shown to improve in accuracy and consistency in responses \cite{hagström2023effectscalingretrievalaugmentation}.  
Debates are well-suited for LLMs and RAG systems because they require strong argumentative skills, along with the ability to recall factual knowledge. 
From these studies, LLMs enhanced with RAG systems are a strong option for use in debates.
Furthermore, studies have shown that LLMs can surpass human performance in debate evaluations \cite{liu2024empiricalanalysislargelanguage}.
\section{Methodology}

We develop a debate system that operates in both the response and evaluation RAD sub-tasks using large language models.
The response sub-task generates an appropriate response to an ongoing debate.
The evaluation sub-task assesses the quality of the given response in the context of the debate.
The interfaces for the debate system adhere to the GenIRSim API, which is used to simulate and evaluate debates \cite{kiesel2024will}.

\begin{table}[htbp]
    \centering
    \caption{Models used for RAD experiments via OpenRouter.}
    \label{tab:llm_models}
    \begin{tabular}{@{}l c r r r@{}}        
        \toprule
        \textbf{Model Name} & \textbf{Release Date} & \textbf{Context} & \textbf{Input} & \textbf{Output} \\
                            &                       & \textbf{(tokens)}     & \textbf{(\$/M)} & \textbf{(\$/M)} \\
        \midrule
        anthropic/claude\_opus-4 & 2025-05-22 & 200,000 & 15 & 75 \\
        anthropic/claude\_sonnet-4 & 2025-05-22 & 200,000 & 3 & 15 \\
        google/gemini-2.5-flash-preview-05-20 & 2025-05-20 & 1,048,576 & 0.15 & 0.60 \\
        google/gemini-2.5-pro-preview & 2025-05-07 & 1,048,576 & 1.25 & 10 \\
        openai/gpt-4.1 & 2025-04-14 & 1,047,576 & 2 & 8 \\
        openai/gpt-4o & 2024-05-13 & 128,000 & 2.50 & 10 \\
        \bottomrule
    \end{tabular}
\end{table}

In exploratory tests, we generated a set of six random topics from the ClaimRev corpus, simulating a debate with a maximum of three turns per model, resulting in six simulations per model for a total of 18 utterances. 
We then fed each model its previous simulations to obtain evaluation scores.

\begin{figure}[p]
    \centering
    \input{sections/33_listing_prompt_debate}
    \caption{
        The prompt was used to generate a debate response in the first sub-task.
        Evidence from the claims database is formatted as yaml at the start of the prompt, while the context of the entire argument is serialized at the end of the prompt.
    }
    \label{fig:prompt_debate}
\end{figure}

\begin{figure}[p]
    \centering
    \include{sections/34_listing_prompt_evaluation}
    \caption{
        The prompt is used to generate all measures of the requested debate.
        The call to the LLM leverages structured output through the chat completion API provided by OpenRouter.
    }
    \label{fig:prompt_evaluation}
\end{figure}

\subsection{Debate Response}

We made a total of six submissions to response generation (sub-task 1) selected from OpenRouter, a platform that provides unified access to an extensive catalog of AI models.
We selected two models each from OpenAI, Anthropic, and Google, as listed in Table~\ref {tab:llm_models}.
We generally select recent flagship models with large context windows.

The debate response has a retrieval and generation phase.
In the retrieval phase, we obtain relevant documents from the argument search system.
The search system utilizes Elasticsearch and Stella embeddings, provided by the competition committee in the baseline system, to retrieve documents relevant to the debate topic \cite{zhang2024jasper}. 
We retrieve the top 10 documents from the search system to use as context.
The generation phase prompts an LLM, shown in Figure \ref{fig:prompt_debate}, with the retrieved documents and the entire state of the debate.
We then respond to the debate request using the interface described in Figure \ref{fig:debate_inteface}.

\begin{figure}[htbp]
    \centering
    \input{sections/31_listing_debate_response}
    \caption{
        The function signature of the responding endpoint the GenIRSim system expects for a valid simulation.
        The messages are passed in OpenAI-compatible chat completion format, with the roles of "user" and "assistant".
    }
    \label{fig:debate_inteface}
\end{figure}

\subsection{Debate Evaluation}

The evaluation of the debate response is handled in a single-shot manner.
Similarly to response generation (sub-task 1) described in Section 3.1, we made a total of six submissions to response evaluation using the models listed in table~\ref{tab:llm_models}.

We prompt the LLM, shown in Figure \ref{fig:prompt_evaluation}, with the most recent responses and the topic of the debate.
We utilize the structured generation interface to generate scores for quantity, quality, manner, and relation.
We are required to expose four separate endpoints for the evaluation measures.
To limit the total number of LLM calls, we evaluate all four measures in a single call.
The resulting call is memoized to avoid repeated calls for the same debate input.

\begin{figure}[htbp]
    \centering
    \input{sections/32_listing_debate_evaluation}
    \caption{
        The function interface for the evaluation portion of the GenIRSim API.
        A separate request is made for quantity, quality, manner, and relation.
    }
    \label{fig:evaluate_inteface}
\end{figure}

\subsection{System Submission}

We submitted to TIRA \cite{froebe:2023}.
Due to the nature of cloud-based LLMs, we are unable to evaluate the system locally.
We opt to submit a thin proxy client to TIRA that calls our self-hosted application service.
Without this, we would need some mechanism to bring the API key into the TIRA environment, which is not possible without exposing the key outside of a trusted environment.
The application service handles RAD requests, logging, and remote LLM calls via OpenRouter, which is exposed over the internet.

\section{Results}

\subsection{Internal Validation}

In our simulation, which runs over six randomly selected debate topics, there were three turns per debate, resulting in 18 system utterances per model. 
From our simulations, Anthropic's Claude Opus 4 model responded the most verbosely, with a mean of 56.5 words per utterance, even peaking at the required maximum of 60 words. 
It also varied the most in words per utterance, with a minimum response of 25 words and the highest standard deviation of 8.4244. 
The sibling Sonnet 4 model also peaked at the required 60-word limit but varied less, with a standard deviation of 3.8348, and was centered around the mean words per utterance at 53.0. 
The Google Gemini models varied in response length, with standard deviations ranging from 6.3 to 6.6. 
Both models had similar maximum and minimum and stayed within the 60-word limit.
Finally, the OpenAI GPT-4 variant models showed the highest consistency, with the lowest standard deviations ranging from 2.2 to 2.7. 
Although they were the most consistent, GPT-4.1 had the second highest mean, 54.4444 words per utterance, which ranged in values between 51 and 59. 
GPT-4o performed with the third-fewest words per utterance.

\begin{table}[htbp]
    \centering
    \caption{Descriptive statistics for words per system utterance by model.}
    \label{tab:model_word_per_utterance_descriptive_stats}
    \begin{tabular}{@{}lcrrrrrr@{}}
        \toprule
        \textbf{Model Name} & \textbf{Release Date} & \textbf{Count} & \textbf{Mean} & \textbf{Median} & \textbf{Std} & \textbf{Min} & \textbf{Max} \\
        \midrule
        touché committee/baseline                & ---        & 422 & 23.4444 & 20.0 & 14.2590 &  7 & 54 \\
        anthropic/claude\_opus-4                 & 2025-05-22 & 1017 & 56.5000 & 59.0 & 8.4244 & 25 & 63 \\
        anthropic/claude\_sonnet-4               & 2025-05-22 & 954 & 53.0000 & 52.0 & 3.8348 & 46 & 61 \\
        google/gemini-2.5-flash-preview-05-20    & 2025-05-20 & 799 & 44.3889 & 43.0 & 6.3165 & 34 & 56 \\
        google/gemini-2.5-pro-preview            & 2025-05-07 & 812 & 45.1111 & 48.0 & 6.5878 & 32 & 54 \\
        openai/gpt-4.1                           & 2025-04-14 & 980 & 54.4444 & 55.0 & 2.2022 & 51 & 59 \\
        openai/gpt-4o                            & 2024-05-13 & 927 & 51.5000 & 52.0 & 2.6624 & 46 & 56 \\
        \bottomrule
    \end{tabular}
\end{table}


For the evaluations, we ran the toy simulation topics through the baseline simulation image provided by the committee.
The toy simulation consists of two topics: "Television is bad" and "Television is good."
For every model, we evaluated the baseline simulations of these debates (two topics, one debate per topic with three messages each) four times. 
We used the baseline image for the simulations to ensure a consistent set of debates for evaluating our results.

In the evaluations, the Anthropic models were the most strict for overall score, with Opus 4 and Sonnet 4 coming in at 0.2773 and 0.2681, respectively.
The standard deviations of both models were less than 0.2600, which is significantly more consistent than those of the other models.
Google's Gemini 2.5 Flash Preview gave the highest overall score of 0.4068, while the Pro Preview gave an overall evaluation of 0.3527. 
OpenAI's GPT-4o gave an overall average of 0.3870, and GPT-4.1 was at 0.3170. 
Interestingly, these are overwhelmingly "no" scores overall. 
However, these averages are weighed down by the models, all of which agree that the Relation, Quantity, and Quality metrics of the simulation responses were low. 
The manner was the only "good" metric evaluated by the models, where each came in at a "yes".

\begin{table}[h]
\centering
\caption{Model evaluations across metrics, including an overall average of means.}
\label{tab:model_evaluation_overall_avg}
\begin{tabular}{|l|r|cccc|}
\hline
\textbf{Model} & \textbf{Overall Avg.} & \textbf{Manner} & \textbf{Quality} & \textbf{Quantity} & \textbf{Relation} \\
\hline
google/gemini-2.5-flash & \textbf{0.407} & 0.694 $\pm$ 0.148 & 0.350 $\pm$ 0.220 & 0.315 $\pm$ 0.184 & 0.269 $\pm$ 0.314 \\
openai/gpt-4o & 0.387 & 0.565 $\pm$ 0.127 & \textbf{0.363 $\pm$ 0.092} & \textbf{0.319 $\pm$ 0.076} & \textbf{0.302 $\pm$ 0.194} \\
google/gemini-2.5-pro & 0.353 & \textbf{0.806 $\pm$ 0.139} & 0.215 $\pm$ 0.212 & 0.167 $\pm$ 0.118 & 0.223 $\pm$ 0.315 \\
openai/gpt-4.1 & 0.317 & 0.635 $\pm$ 0.093 & 0.247 $\pm$ 0.102 & 0.219 $\pm$ 0.099 & 0.167 $\pm$ 0.178 \\
anthropic/claude\_opus-4 & 0.277 & 0.519 $\pm$ 0.153 & 0.243 $\pm$ 0.085 & 0.171 $\pm$ 0.059 & 0.177 $\pm$ 0.254 \\
anthropic/claude\_sonnet-4 & 0.268 & 0.500 $\pm$ 0.186 & 0.238 $\pm$ 0.079 & 0.215 $\pm$ 0.048 & 0.120 $\pm$ 0.180 \\
\hline
\end{tabular}
\end{table}

\subsection{Leaderboard Evaluation}

In addition to our internal evaluation, we submitted multiple runs to the official Touché 2025 shared task leaderboard. 
The results for both sub-tasks are shown in Table~\ref{tab:touche25_subtask1_results} and Table~\ref{tab:touche25_subtask2_results}. 
Sub-task 1 evaluates the proportion of responses that fulfill each of the Gricean maxims, while sub-task 2 is a classification task reporting precision, recall, and F1 for each maxim. 
Our GPT-4.1 and Gemini-2.5 runs ranked among the top submissions in sub-task 1 and showed competitive results in sub-task 2.

\begin{table}[htbp]
\centering
    \caption{Official Touché 2025 results for Sub-Task 1: Proportion of responses fulfilling each of the Gricean maxims (Quantity, Quality, Relation, Manner) across submitted runs.}
    \label{tab:touche25_subtask1_results}
\begin{tabular}{|l|l|c|c|c|c|c|}
\hline
\textbf{Team} & \textbf{Run} & \textbf{Score (AVG)} & \textbf{Quantity} & \textbf{Quality} & \textbf{Relation} & \textbf{Manner} \\
\hline
Baseline   & baseline            & 0.62          & 0.35          & \textbf{1.00} & 0.32 & 0.80 \\
\hline
DS@GT & gpt-4.1             & \textbf{0.70} & \textbf{0.95} & 0.17 & 0.82 & \textbf{0.84} \\
Team DS@GT & gemini-2.5-pro          & 0.65          & 0.94          & 0.26 & 0.74 & 0.67 \\
DS@GT & gemini-2.5-flash    & 0.50          & 0.70          & 0.07 & 0.80 & 0.41 \\
DS@GT & claude-opus-4       & 0.42          & 0.41          & 0.31 & \textbf{0.87} & 0.09 \\
DS@GT & gpt-4o              & 0.42          & 0.20          & 0.02 & 0.86 & 0.58 \\
DS@GT & claude-sonnet-4     & 0.38          & 0.35          & 0.05 & \textbf{0.94} & 0.17 \\
\hline
\end{tabular}
\end{table}

\begin{table}[htbp]
\centering
    \caption{Official Touché 2025 results for Sub-Task 2: Precision (P), Recall (R), and F1-score for each maxim classification task.}
    \label{tab:touche25_subtask2_results}
\resizebox{\textwidth}{!}{
\begin{tabular}{|l|l|c|ccc|ccc|ccc|ccc|}
\hline
\textbf{Team} & \textbf{Run} & \textbf{Score (F1)} & \multicolumn{3}{c|}{\textbf{Quantity}} & \multicolumn{3}{c|}{\textbf{Quality}} & \multicolumn{3}{c|}{\textbf{Relation}} & \multicolumn{3}{c|}{\textbf{Manner}} \\
             &              &                      & \textbf{P} & \textbf{R} & \textbf{F1} & \textbf{P} & \textbf{R} & \textbf{F1} & \textbf{P} & \textbf{R} & \textbf{F1} & \textbf{P} & \textbf{R} & \textbf{F1} \\
\hline
Baseline       & 1-baseline         & \textbf{0.67} & 0.57 & \textbf{1.00} & 0.73 & 0.24 & \textbf{1.00} & 0.38 & 0.78 & \textbf{1.00} & 0.87 & 0.52 & \textbf{1.00} & \textbf{0.68} \\
 DS@GT     & gemini-2.5-flash   & 0.64          & 0.59 & 0.86          & 0.70 & 0.18 & 0.66          & 0.29 & 0.81 & 0.99          & 0.89 & 0.52 & 0.99          & \textbf{0.68} \\
 DS@GT     & gpt-4o             & 0.64          & 0.59 & 0.88          & 0.71 & 0.17 & 0.63          & 0.27 & 0.82 & 0.99          & 0.89 & 0.52 & 0.97          & 0.67 \\
 DS@GT     & gpt-4.1            & 0.62          & 0.58 & 0.75          & 0.65 & 0.15 & 0.52          & 0.24 & 0.82 & 0.98          & \textbf{0.90} & 0.52 & 0.99          & \textbf{0.68} \\
 DS@GT     & gemini-2.5-pro     & 0.62          & 0.59 & 0.67          & 0.63 & 0.17 & 0.52          & 0.25 & 0.84 & 0.97          & \textbf{0.90} & 0.52 & 0.98          & \textbf{0.68} \\
 DS@GT     & claude-sonnet-4    & 0.56          & 0.43 & 0.49          & 0.49 & 0.15 & 0.36          & 0.21 & 0.83 & 0.92          & 0.88 & 0.51 & 0.93          & 0.66 \\
 DS@GT     & claude-opus-4      & 0.51          & 0.49 & 0.21          & 0.29 & 0.16 & 0.31          & 0.21 & 0.85 & 0.90          & 0.88 & 0.51 & 0.92          & 0.66 \\
\hline
\end{tabular}
}
\end{table}

\subsection{Projected Costs for Evaluation}

We provide statistics on debate generation and evaluation that take place after the task submission deadline.
The spend, tokens, and request statistics are derived from the OpenRouter dashboard and may include a mix of official evaluation requests and development requests.
The team spent \$151.63 in total credits across all models for the duration of the competition.

Table~\ref{tab:gen_task_analysis} covers the generation task on June 24th, 2025.
There are approximately 200 requests per submission, with a high ratio of input tokens to output tokens at 95\%.
The Gemini Pro model is handled differently because "thinking" tokens are treated as output tokens.
This ratio makes sense since we are generating a short 60-word response.

\begin{table}[htbp]
\centering
\caption{Generation Task OpenRouter Usage Summary on 2025-06-24}
\label{tab:gen_task_analysis}
\begin{tabular}{lrrrr}
\toprule
\textbf{Model Name} & \textbf{Spend (\$)} & \textbf{Tokens (K)} & \textbf{Requests} & \textbf{Input \%} \\
\midrule
google/gemini-2.5-flash-preview-05-20 & \$0.045 & 270 & 202 & 95.9\% \\
openai/gpt-4.1-2025-04-14             & \$0.662 & 291 & 212 & 95.5\% \\
openai/gpt-4o                         & \$0.819 & 289 & 211 & 95.5\% \\
anthropic/claude-4-sonnet-20250522    & \$1.09  & 306 & 202 & 95.4\% \\
anthropic/claude-4-opus-20250522      & \$5.21  & 289 & 190 & 94.8\% \\
google/gemini-2.5-pro-preview-06-05   & \$3.17  & 547 & 201 & 48.1\% \\
\bottomrule
\end{tabular}
\end{table}

We perform a similar analysis for the evaluation task in Table~\ref{tab:eval_task_analysis}.
Each evaluator is run against the generated results from eight debate subsystems across 100 topics, each with five turns.
A complete run of evaluation tasks against Gemini 2.5 Flash costs about \$1 USD.
We note that some of the models are not necessarily cost-effective, as this task requires significant context.

\begin{table}[htbp]
\centering
\caption{Evaluation Task OpenRouter Usage Summary on 2025-06-25}
\label{tab:eval_task_analysis}
\begin{tabular}{lrrrr}
\toprule
\textbf{Model Name} & \textbf{Spend (\$)} & \textbf{Tokens (K)} & \textbf{Requests} & \textbf{Input \%} \\
\midrule
google/gemini-2.5-flash-preview-05-20 & \$1.13   & 4,580 & 4,930 & 78.5\% \\
openai/gpt-4.1-2025-04-14             & \$0.372  & 117   & 119   & 80.3\% \\
openai/gpt-4o                         & \$0.188  & 51    & 54    & 84.1\% \\
anthropic/claude-4-sonnet-20250522    & \$0.843  & 136   & 148   & 73.3\% \\
anthropic/claude-4-opus-20250522      & \$12.10  & 427   & 493   & 77.8\% \\
google/gemini-2.5-pro-preview-06-05   & \$6.64   & 894   & 366   & 29.4\% \\
\bottomrule
\end{tabular}
\end{table}

We then project what the cost of each of the models looks like on the complete evaluation task in Table~\ref{tab:projected_cost}.
We pad the total number of requests by 1000 for a total of 5000 requests (i.e., eight subsystems, 100 topics, five rounds) to account for development and debugging.
The Claude Opus and Gemini 2.5 Pro models are significantly more expensive than the other models due to their extended thinking token budgets.
Due to the cost projections, Claude Opus was not run in the final evaluation.

\begin{table}[htbp]
\centering
\caption{Projected cost for the Evaluation Task (5,000 Requests), using the 79\% input / 21\% output ratio from the Gemini Flash evaluation run.}
\label{tab:projected_cost}
\begin{tabular}{lr}
\toprule
\textbf{Model Name} & \textbf{Estimated Total Cost} \\
\midrule
google/gemini-2.5-flash-preview-05-20 & \$1.13 \\
openai/gpt-4.1-2025-04-14             & \$15.86 \\
openai/gpt-4o                         & \$19.92 \\
anthropic/claude-4-sonnet-20250522    & \$26.88 \\
google/gemini-2.5-pro-preview-06-05   & \$81.41 \\
anthropic/claude-4-opus-20250522      & \$134.40 \\
\bottomrule
\end{tabular}
\end{table}

\section{Discussion}

During the early stages of this project, we prototyped our approaches to various tools that we finalized in the submission.
We used TensorZero as a starting point for integrating models from multiple providers and exploring structured prompting, episode generation, and call logging.
However, we found that maintaining our system incurred overhead, and the conveniences afforded by using a single API key vastly outweighed any benefits from an API gateway. We also implemented a debate manager that enabled us to run debates using our model through a web interface written in Streamlit. The debate manager was structured using the pytransitions library, which allowed us to define a strict state machine for the debate process.

The final submission is vastly simplified by GenIRSim, which provided the debate manager and episode generation. However, figuring out how to use the GenIRSim API required reading the source code and the API documentation.

We find it difficult to evaluate both the quantitative and qualitative aspects of the debate during development. Although the debate manager provides a way to run the debates and adjust our prompts, it is challenging to apply empirical methods to improve the debate process iteratively. The task at hand is complex and requires an experimental setting with experts to evaluate the quality of the debate.

\section{Future Work}

We aim to explore finer-grained control over the agent's behavior.
It would be beneficial to modify the agent's behavior to more accurately simulate diverse perspectives and personalities and to encourage participants to strategize in ways that capitalize on the strengths and exploit the weaknesses in discourse.
Behaviors can be implemented via controllable parameters that govern specific strategy types and explored by having several sets of behaviors, which, while not necessarily realistic, are easy to examine under controlled conditions.
Strategies could include always deflecting, always attacking, or always conceding.

Another area of future work is formulating a more robust evaluation framework.
Although we have a basic evaluation system, having more time to explore the results of the evaluation between different models would be interesting in the context of a leaderboard.
These results have to be grounded in human evaluation, but an ELO system could be used to rank models based on their performance in debates.

We also did not find time to explore retrieval strategies.
There are more complex strategies than top-k dense retrieval that could improve the quality of responses.
\section{Conclusions}
As seen throughout this task, LLMs excel at summarizing queried arguments, but they are notably verbose in their responses compared to the baseline debating system. Additionally, the LLMs were strict in evaluating the baseline system's utterances, so it will be interesting to see the results of the true positives and negatives when compared to the baseline evaluators. The models were pessimistic on the same metrics, but in different magnitudes. Therefore, exploring evaluations on a much larger set of simulations would be worthwhile.
The accompanying source code for this paper is located at \url{https://github.com/dsgt-arc/touche-2025-rad}.
\section*{Acknowledgments}

We thank the Data Science at Georgia Tech (DS@GT) CLEF competition group for their support.

\section*{Declaration on Generative AI}

 During the preparation of this work, the authors used Gemini in order to: grammar and spelling check, formatting assistance, peer review simulation. 
 After using these tools/services, the authors reviewed and edited the content as needed and each take full responsibility for the publication’s content. 
\bibliography{main}
\end{document}